\documentclass[
 reprint,
superscriptaddress,
 amsmath,amssymb,
 aps,
pra]{revtex4-2}

\usepackage{graphicx}

\usepackage{dcolumn}
\usepackage{bm}

\usepackage{hyperref}
\usepackage{braket}
\usepackage{physics}
\usepackage{siunitx}

\usepackage{comment}
\usepackage{ amssymb }
\usepackage[cmyk]{xcolor}
\usepackage{booktabs}

\hypersetup{hidelinks}

\begin{document}

\title{Frequency Fluctuations in Nanomechanical Resonators due to Quantum Defects}

\renewcommand{\andname}{\ignorespaces}

\author{M. P. Maksymowych}
\thanks{These authors contributed equally}
\affiliation{Department of Applied Physics and Ginzton Laboratory, Stanford University; Stanford CA, 94305, USA.}

\author{M. Yuksel}
\thanks{These authors contributed equally}
\affiliation{Department of Applied Physics, California Institute of Technology; Pasadena CA, 91125, USA.}

\author{O. A. Hitchcock}
\affiliation{Department of Physics, Stanford University; Stanford CA, 94305, USA.}

\author{N. R. Lee}
\affiliation{Department of Applied Physics and Ginzton Laboratory, Stanford University; Stanford CA, 94305, USA.}

\author{F. M. Mayor}
\affiliation{Department of Applied Physics and Ginzton Laboratory, Stanford University; Stanford CA, 94305, USA.}

\author{W. Jiang}
\affiliation{Department of Applied Physics and Ginzton Laboratory, Stanford University; Stanford CA, 94305, USA.}

\author{M. L. Roukes}
\affiliation{Department of Applied Physics, California Institute of Technology; Pasadena CA, 91125, USA.}
\affiliation{Department of Physics, Department of Bioengineering and the Kavli Nanoscience Institute, California Institute of Technology; Pasadena CA, 91125, USA.}

\author{A. H. Safavi-Naeini}
\email{safavi@stanford.edu}
\affiliation{Department of Applied Physics and Ginzton Laboratory, Stanford University; Stanford CA, 94305, USA.}

\begin{abstract} 

Nanomechanical resonators promise diverse applications ranging from mass spectrometry to quantum information processing, requiring long phonon lifetimes and frequency stability. Although two-level system (TLS) defects govern dissipation at millikelvin temperatures, the nature of frequency fluctuations remains poorly understood. In nanoscale devices, where acoustic fields are confined to sub-wavelength volumes, strong coupling to individual TLS should dominate over weak coupling to defect ensembles. In this work, we monitor fast frequency fluctuations of phononic crystal nanomechanical resonators, while varying temperature ($10$~mK$-1$~K), drive power ($10^2-10^5$ phonons), and the phononic band structure. We consistently observe random telegraph signals (RTS) which we attribute to state transitions of individual TLS. The frequency noise is well-explained by mechanical coupling to individual far off-resonant TLS, which are either thermally excited or strongly coupled to thermal fluctuators. Understanding this fundamental decoherence process, particularly its RTS structure, opens a clear path towards noise suppression for quantum and sensing applications.

\end{abstract}

\date{\today}

\maketitle

Mechanical resonators have emerged as a crucial platform for quantum applications, spanning sensing, computation, and communication \cite{andrews2014bidirectional, jiang2023optically, meesala2024non, pechal2018, hann2019hardware, chamberland2022building}. These applications demand exquisite coherence, driving research towards devices with higher quality factors and reduced dephasing. Two approaches have achieved remarkable success in realizing long lifetimes: phononic crystals and bulk acoustic wave (BAW) devices \cite{maccabe2020nano, kharel2018ultra}. For many sensing applications, such as detecting minute masses or forces, nanoscale devices~\cite{naik2009towards,neumann2024nanomechanical, sage2018single, hanay2012single, hanay2015inertial, maksymowych2019optomechanical} including those based on phononic crystals present a promising platform. The small size of these resonators leads to high responsivity. Moreover, recent advances in integrating these devices with quantum superconducting circuits have enabled the generation of nonclassical and entangled states, and new types of quantum measurement~\cite{arrangoiz2019resolving, cleland2024studying, wollack2022quantum}. 

Despite recent advances, current devices are limited by decoherence, which affects all quantum and sensing applications discussed above ~\cite{jiang2023optically, meesala2024non, wollack2021loss, yu2014phononic, chen2024phonon, bozkurt2024mechanical}. At millikelvin temperatures, where many quantum experiments are conducted, most sources of dissipation and noise such as nonlinear phonon scattering \cite{atalaya2016nonlinear, srivastava2022physics}, adsorption-desorption \cite{yang2011surface}, defect diffusion \cite{fong2012frequency, atalaya2011diffusion}, and thermomechanical noise \cite{wang2023beating, ekinci2004ultimate, cleland2002noise, sansa2016frequency} become irrelevant. Ultra-low temperatures suppress thermally activated processes, while high vacuum conditions prevent adsorption-desorption events that could perturb the resonator. A remaining source of loss at millikelvin temperatures is attributed to weak interactions with 
two-level system (TLS) ensembles, often described by the standard tunneling model (STM) ~\cite{phillips1987two, behunin2016dimensional, gao2008physics, muller2019towards,emser2024thinfilmquartzhighcoherencepiezoelectric}, much like macroscopic surface acoustic wave \cite{gruenke2024surface,andersson2021acoustic, manenti2016surface} and superconducting ~\cite{martinis2005decoherence,gao2008experimental,gao2008physics, burnett2016analysis, lucas2023quantum, muller2019towards, mcrae2020materials} devices. However, when fields are confined to a nanoscale volume, strong interactions with few TLS should dominate over ensemble effects, challenging our understanding of dissipation in cryogenic nanomechanical resonators. Moreover, recent experiments have revealed that significant dephasing persists even in carefully engineered nanomechanical devices at millikelvin temperatures ~\cite{ maccabe2020nano, kalaee2019quantum, bozkurt2023quantum, cleland2024studying}. These large frequency fluctuations have not been studied in detail and their sources are unknown, necessitating a thorough investigation.

\begin{figure*}
    \centering
        \includegraphics[width=0.8\linewidth]{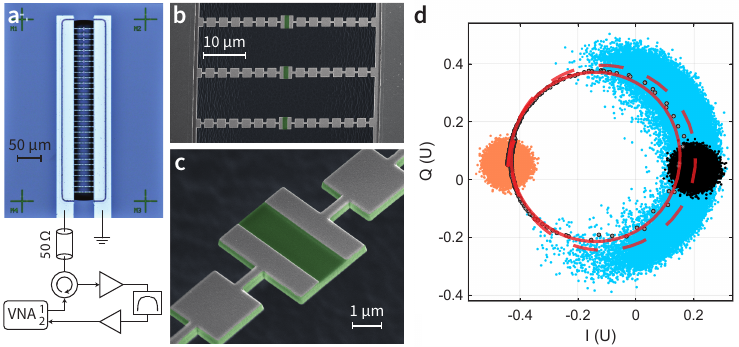}
        \caption{\textbf{Spectral diffusion of a phononic crystal.} (a) Optical microscope image of a phononic crystal resonator (PCR) array with aluminum contact pads connected to a \SI{50}{\ohm} transmission line and a vector network analyzer (VNA) via a cryogenic circulator. (b) Colorized scanning electron micrograph (SEM) of frequency multiplexed PCRs (Green: LiNbO$_3$, Grey: Al). (c) A 45 degree colorized SEM of a single PCR. (d) An IQ space plot of $S_{11}$ data sampled at different drive frequencies (grey dots) near a \SI{751.7}{\MHz} PCR mode (PCR 1) at \SI{10}{\milli\kelvin} and fitted to equation \ref{eq:S11} (solid red line) assuming no fluctuations. The $S_{11}$ fluctuations acquired off-resonance (orange), on-resonance (blue), and the inferred on-resonance detection noise (black) are shown. The resonator locus with the dephasing-corrected relaxation rate $\kappa_i'$ is shown as a red dashed line. } 
    \label{figure1}
\end{figure*}

In this work, we study decoherence in nanopatterned phononic crystal resonators (PCRs) comprised of thin-film lithium niobate (LN). Our investigation focuses on observing and characterizing frequency fluctuations, particularly their frequency, power, and temperature dependence. Our main results are: i) The spectral diffusion in our devices is dominated by a random telegraph signal (RTS), likely arising from state changes of a single TLS. ii) Frequency fluctuations appear to reduce 100-fold by going from 10 mK to 1 K, which we show is caused by the TLS switch rate surpassing the cavity bandwidth.
iii) Our measurements reveal that while intrinsic (high drive power) quality factors can be enhanced to above $\sim$10 million using acoustic bandgaps, spectral diffusion persists and is insensitive to the phononic band structure. iv) We observe that dephasing continues even with $10^5$ intra-cavity phonons, suggesting it's caused by a far off-resonant TLS rather than resonant TLS. Borrowing notions from the generalized tunneling model (GTM)~\cite{faoroIoffe2015interacting}, the frequency noise can be explained by strong coupling of a single high energy TLS to a two-level fluctuator (TLF). These findings show a dominant mechanism of decoherence in nanomechanical systems and reveal avenues for circumventing dephasing in applications.

\section*{Spectral Diffusion of Nanomechanical Resonator Arrays} 
We investigate on-chip arrays of mechanical resonators based on phononic crystals fabricated from \SI{250}{\nm} thick X-cut LN ~\cite{lee2023strong}.  Each PCR consists of a one-dimensional periodic array of cells supporting a complete phononic bandgap. A defect at the center of this crystal breaks the periodicity and supports a mechanical resonance. Two aluminum (Al) electrodes deposited on top of the PCR enable piezoelectric transduction of the fundamental shear mode (see Fig.~\ref{figure1}a-c). Device fabrication is detailed in the Methods. Instead of connecting the electrodes to a superconducting qubit~\cite{arrangoiz2019resolving,lee2023strong}, we probe devices directly via microwave-frequency reflectometry~\cite{wollack2021loss}. We address arrays of devices coupled to a single RF line, enabling parallel measurements of multiple resonators. In this work, we focus primarily on an array of 31 resonators with frequencies ranging from \SI{450}{\MHz} to 1 GHz. Near each resonance, the reflection $S$-parameter is given by:
\begin{align}
	S_{11}(\omega) = I+iQ = S_\mathrm{bg}\times\Bigg(1-e^{i\phi}\frac{\kappa_e/\cos(\phi)}{i\Delta(t)+\frac{\kappa(t)}{2}}\Bigg) \label{eq:S11}
\end{align}
where $\Delta(t) = \omega-\omega_0(t)$ is the detuning, $\omega_0(t)=\bar{\omega}_0+\delta\omega_0(t)$ is the resonant frequency, $\kappa(t)=\kappa_e+\kappa_i(t)$ is the total loss rate, $\kappa_e$ is the extrinsic coupling rate, $\kappa_i(t)=\bar{\kappa}_i+\delta\kappa_i(t)$ is the intrinsic loss rate, $S_\mathrm{bg}$ is a microwave background fit, and $\phi$ is a cavity asymmetry parameter which partially accounts for reflections in the measurement line ~\cite{khalil2012analysis}. The time-dependence of the frequency and linewidth captures their fluctuations over time. We can assume the above form of a time-dependent $S$-parameter so long as the fluctuations are slow compared to the resonance frequency, linewidth, and the time required to determine the $S$-parameter~\cite{SI}.

\begin{figure*}
    \centering
        \includegraphics[width=\linewidth]{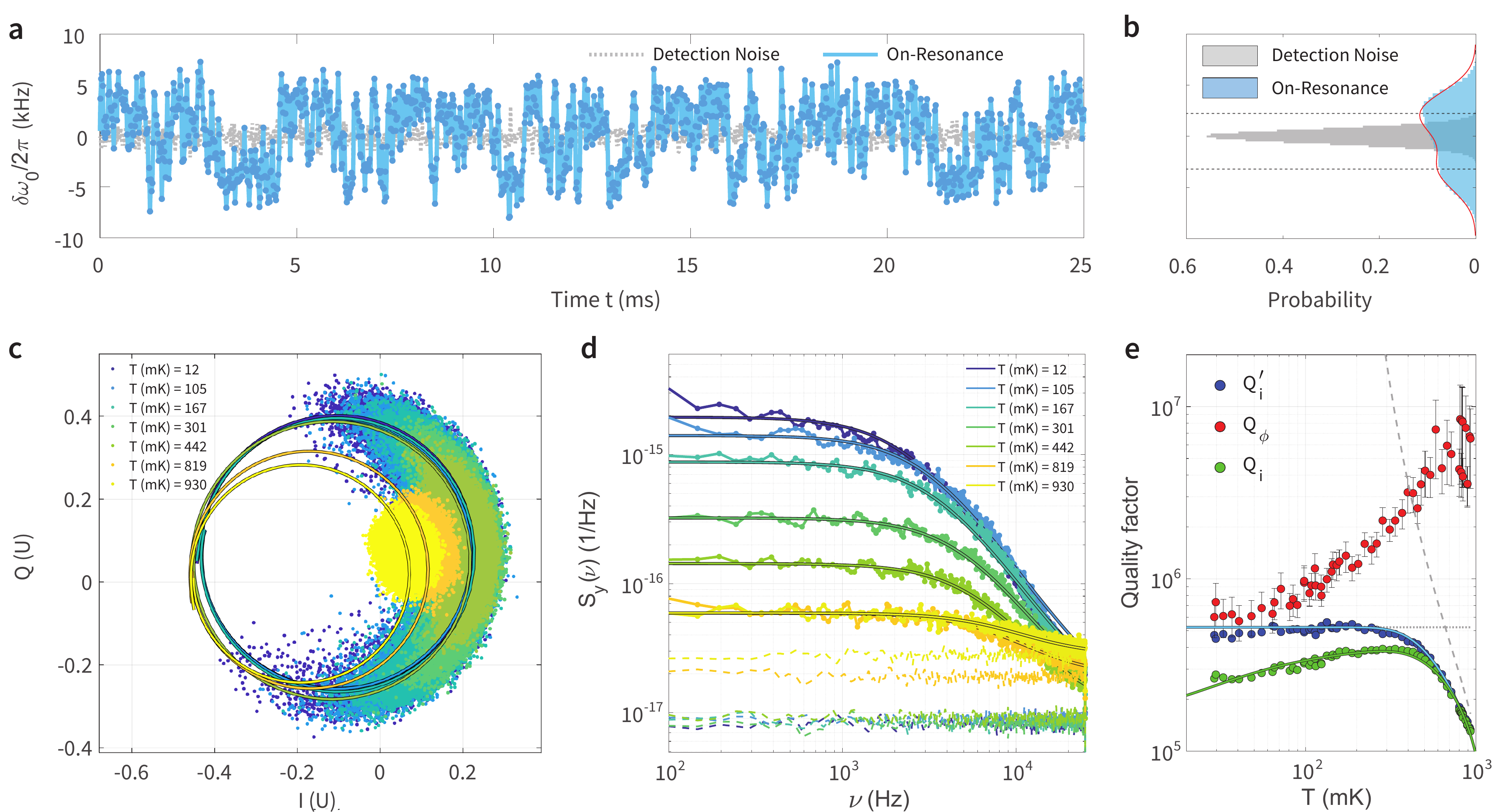}
        \caption{\textbf{Telegraph frequency jumps at millikelvin temperatures.} (a) The frequency fluctuations of PCR~1 while probed on resonance $\delta\omega_0(t)/2\pi$ (blue), and the associated detection noise (grey) versus time ($n \approx 5\times 10^3, T \approx \SI{10}{\milli\kelvin}$). The corresponding probability distributions are shown in (b), where the on-resonance data is fitted to a two-mean Gaussian (red). (c) The raw $S_{11}(t)$ fluctuation data plotted at different temperatures with fixed $n\approx 1\times 10^4$ for PCR 1 (dots) and the corresponding dephasing-corrected loci (lines). (d) The power spectral density (PSD) of the fractional frequency fluctuations $S_y(\nu)$ for the data in (c), with fits to equation~\ref{eq:Sy} displayed as solid lines and detection noise as dashed lines.(e) The temperature dependence of $Q_i$, $Q_\phi$, and $Q_i'$ in green, red and blue, respectively, for PCR 3 probed with $n=3\times 10^4$. Dashed grey line: Mattis-Bardeen (MB) losses, Dotted grey line: temperature-insensitive losses, Blue line: temperature-insensitive \& MB loss, Green Line: MB, temperature-insensitive \& resonant TLS loss ~\cite{wollack2021loss} (which is not present in the $Q_i'$ fit; blue line).
        }
    \label{figure2}
\end{figure*}

\begin{figure}
    \centering
        \includegraphics[width=\linewidth]{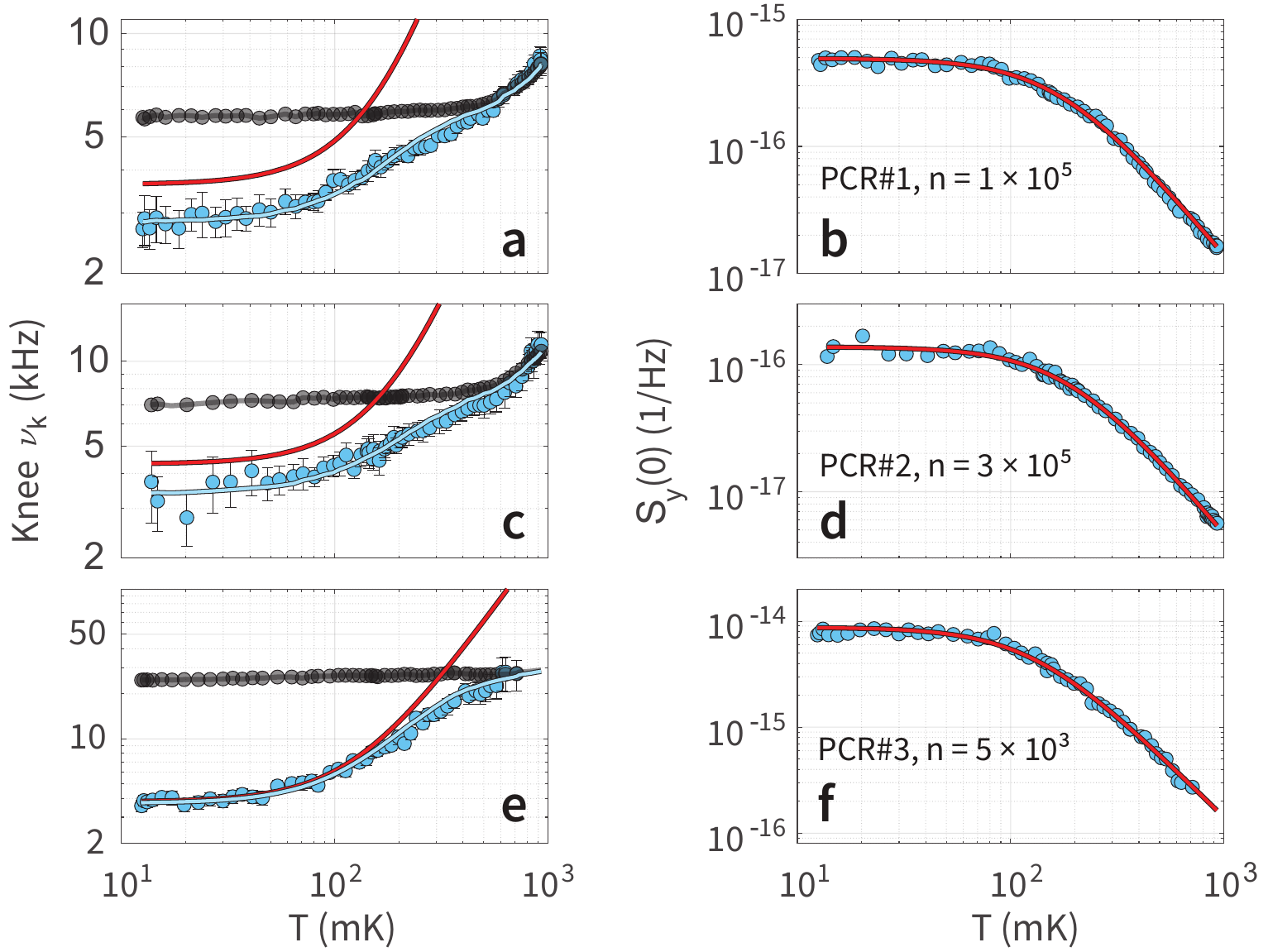}
        \caption{\textbf{Thermally enhanced switching and cavity-based averaging.} Panels (a,b), (c,d) and (e,f) plot the fitted PSD knee and amplitude ($\nu_k=\Gamma/\pi$, $S_y(0)$) as blue circles for PCR 1, 2 and 3, respectively. Each device is probed as function of temperature with fixed cavity phonon number. The cavity knee $\kappa/4\pi$, estimated RTS knee $\hat\Gamma_s/\pi = 2(\alpha +\beta T^2)$ and expected knee $\hat\Gamma/\pi$ of $S_{\text{filt}}(\nu)$ are shown as grey, red and blue lines, respectively, in panels (a), (c) and (e). The values of the switch rate model are: $\alpha = \{1.8, 2.2, 1.9\}$ kHz and $\beta = \{62.2, 60.9, 117.0\} \text{ kHz/K}^{2}$. The $S_y(0)$ shown in panels (b), (d) and (f) is modeled using $\mathcal{A}_0^2/2\hat\Gamma_s$ (red lines) by assuming a temperature-independent $\mathcal{A}=\mathcal{A}_0= \{3.4, 1.93, 14.6\}\times 10^{-6}$ for PCR 1, 2 and 3, respectively.
        }
    \label{figure3}
\end{figure}

To study phonon decoherence under conditions relevant for quantum experiments, the sample chip is thermalized to the mixing plate (MXC) of a dilution refrigerator and cooled to \SI{10}{\milli\kelvin}. We make measurements using a vector network analyzer (VNA; Rohde \& Schwarz ZNB20) outputting a microwave tone from port 1 that reflects off the device into port 2 through a cryogenic circulator (see Fig.~\ref{figure1}a). Fig.~\ref{figure1}d shows data from a \SI{751.7}{\MHz} mode (PCR 1) at $T = \SI{10}{\milli\kelvin}$. The parameters of our PCRs are in Supplementary Table \ref{table:DevParams}. An input transmission line power $P$ at the PCR drives the mode into a coherent state with mean intra-cavity phonon number $n \approx (P/\hbar\bar{\omega}_0) \times 4\kappa_e/(\bar\kappa_i+\kappa_e)^2 \approx 1\times 10^4$. We obtain samples of $S_{11}$ with an integration time of \SI{10}{\ms} at drive frequencies $\omega$ near resonance. We plot these samples as grey dots in Fig.~\ref{figure1}d and fit to the locus described by equation~\ref{eq:S11} (solid red line). Next, we take 100,000 samples of $S_{11}$ with an integration time of \SI{20}{\us} at fixed on- and off-resonance detunings ($\bar{\Delta} = \{0, -25\bar\kappa\}$), plotted as blue and orange dots, respectively. We attribute the spread in the samples of $S_{11}$ taken with the off-resonant drive to detection noise, \textit{e.g.,} noise from the amplifiers. For clarity, we shift this data to the right side of the locus corresponding to small $\Delta$ (black dots) to more easily compare it to the samples from on-resonant driving. The actual samples of $S_{11}$ obtained from on-resonant driving (blue) show a far larger spread which approximately follows the shape of the locus. We attribute this to real parameter noise, \textit{i.e.}, fluctuations in the frequency or linewidth. Frequency fluctuations manifest as excess noise tangential to the locus~\cite{gao2008physics, fong2012frequency, neill2013fluctuations}, much like our data.  

The locus obtained from a drive frequency sweep (Fig.~\ref{figure1}d; solid red) consistently has smaller diameter than the locus that the $S_{11}$ fluctuations (blue) appear to follow. This arises because the integration time of the drive frequency sweep (10 ms) is larger than the characteristic timescale of the frequency fluctuations, $\tau_0$. The solid red locus can therefore be thought of as an average over many loci with slightly different resonant frequencies. From the averaged locus we would infer a linewidth $\kappa_i$ larger than the true intrinsic linewidth $\kappa_i^\prime$, leading to a smaller locus diameter. The samples (blue points) are acquired with a fixed frequency tone and a shorter integration time (\SI{20}{\us}) which leads to less averaging. We therefore use this data to infer $\kappa_i^\prime$ and the spectral linewidth broadening $\kappa_\phi \equiv \kappa_i-\kappa_i'\geq0$ due to frequency fluctuations. To determine $\kappa_i'$, we introduce a locus correction method (LCM):
\begin{enumerate}
    \item We start with locus parameters: $\mathbf{p}=\{\kappa_i',\kappa_e,\bar\omega_0,\phi,S_{bg}\}$, where $\kappa_i'$ is the intrinsic loss rate we aim to determine.
    \item For each measured sample $S_j=I_j+iQ_j$ (blue points), we estimate the corresponding mechanical frequency $\omega_{0,j}$ by minimizing the squared distance to the locus:
    \begin{equation}
        \omega_{0,j} = \text{argmin}_{\omega} |S_j-S_{11} (\omega,\mathbf{p})|^2
    \end{equation}
    \item We calculate the total loss $\mathcal{L}(\kappa_i')$ as the sum of these minimized squared distances:
    \begin{equation}
        \mathcal{L}(\kappa_i') = \sum_j |S_j-S_{11} (\omega_{0,j},\mathbf{p})|^2
    \end{equation}
    \item We determine the optimal $\kappa_i'$ by minimizing this total loss:
    \begin{equation}
        \kappa_i' = \text{argmin}_{\kappa_i'} \mathcal{L}(\kappa_i')
    \end{equation}
\end{enumerate}
The resulting $\kappa_i'$ estimates the intrinsic loss rate, while $\kappa_\phi$ quantifies the spectral broadening from frequency noise and is agnostic to the noise color. The re-sized locus with $\kappa_i'/2\pi\approx \SI{3.5}{\kHz}$ is shown as a dashed red line in Fig.~\ref{figure1}d, and we find $\kappa_\phi/2\pi \equiv (\kappa_i-\kappa_i')/2\pi \approx 1.1$ kHz. This LCM procedure can convert $S_j\to\omega_{0,j}$ and estimate $\{\kappa_i'$,  $\kappa_\phi\}$ for any resonator, provided the frequency fluctuations are slow compared to the integration time used to sample $S_j$ and the measurement is not dominated by detection noise. 

\section*{Temperature Dependence of Phonon Decoherence}
The frequency fluctuations $\delta\omega_0(t)$ of PCR 1 inferred from on-resonance and detection noise data are shown in Fig.~\ref{figure2}a as blue and grey, respectively ($n \approx 5\times 10^3, T \approx \SI{10}{\milli\kelvin}$). The on-resonance fluctuations are dominated by jumps between two frequency-distinct states. This random telegraph signal (RTS) is characterized by a jump amplitude $2\times A/2\pi$ and switch rate $\Gamma_s/2\pi=1/2\tau_0$. The probability distributions of the frequency are in Fig.~\ref{figure2}b, where the on-resonance data is fit to a two-mean Gaussian function yielding a jump amplitude of $2\times A/2\pi=\SI{6.0\pm1.0}{\kHz}$. To more thoroughly investigate the RTS, we tune the temperature via a resistive heater thermalized to the MXC plate and repeat the measurements shown in Fig.~\ref{figure1}d. Fig.~\ref{figure2}c depicts the on-resonance $S_{11}$ fluctuations and LCM re-sized loci versus temperature measured with fixed $n\approx 1\times 10^4$ (detection noise not shown). The power spectral density (PSD) of the fractional frequency fluctuations, $S_y(\nu)$, can be computed from the time series data, $y(t) = \delta\omega_0(t)/\bar{\omega}_0$, via the Welch method. The PSDs corresponding to the data in Fig.~\ref{figure2}c are plotted in Fig.~\ref{figure2}d and fit to a Lorentzian model (solid lines):
\begin{align}
    S_y(\nu)=\frac{\mathcal{A}^2}{2\Gamma}\frac{1}{1+(\pi\nu/\Gamma)^2}+S_0(\nu) \label{eq:Sy}
\end{align}
which includes a frequency-independent detection noise contribution $S_0(\nu)$, shown as a dashed line for each temperature. We parameterize the PSD with the knee frequency $\Gamma/\pi$ and the noise amplitude at zero frequency $S_y(0) = \mathcal{A}^2/2\Gamma$. For PCR~3, Fig.~\ref{figure2}e shows three quality factors: $Q_i\equiv\bar{\omega}_0/\kappa_i$ measured from frequency sweeps with 10 ms integration time, $Q_i'\equiv\bar{\omega}_0/\kappa_i'$ determined via the LCM, and the corresponding dephasing quality factor $Q_\phi\equiv\bar{\omega}_0/\kappa_\phi$. At high temperatures, both $Q_i$ and $Q_i^\prime$ become limited by resistive quasiparticle losses in the electrodes, in agreement with predictions of Mattis-Bardeen theory ~\cite{wollack2021loss}. At lower temperatures, the difference between $Q_i$ and $Q_i^\prime$ becomes larger, which we attribute to an increased role of dephasing in determining the spectral linewidth of the mechanical system. 

The frequency fluctuations shown in Fig.~\ref{figure2} appear to decrease dramatically with temperature, showing more than 10 dB reduction from $\SI{10}{\milli\kelvin}$ to $\SI{1}{\kelvin}$, with an apparent maximum stability between \qtyrange{1}{4}{\K}. The key question we focus on now is whether this reduction reflects a real improvement in phonon coherence, or if it is an artifact of measurement limitations, particularly averaging by the cavity or instrumentation.

\begin{figure*}
    \centering
        \includegraphics[width=1\linewidth]{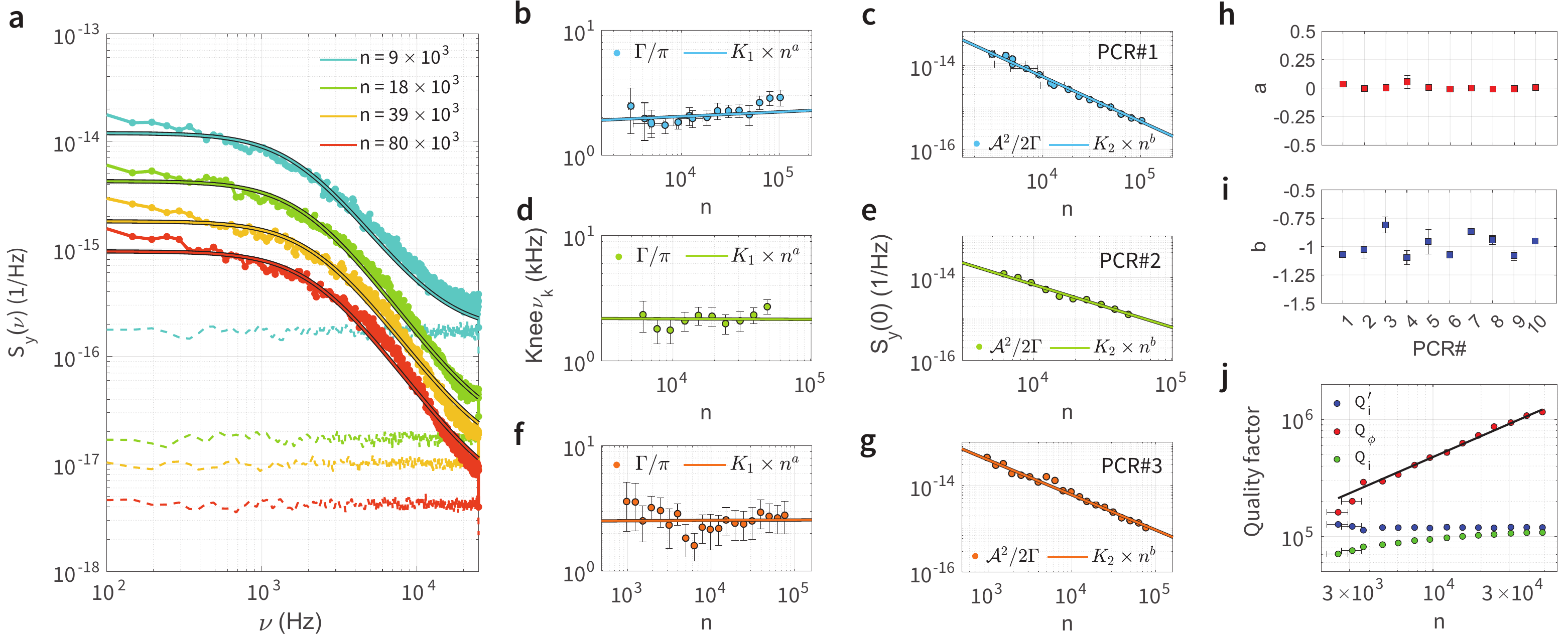}
        \caption{\textbf{Decoherence dependence on coherent drive power.}  (a) The frequency noise power spectral densities $S_y(\nu)$ of PCR 1 at different $n$ (dot-line) and $\SI{10}{\milli\kelvin}$, which are fitted to equation~\ref{eq:Sy} (solid lines). The corresponding detection noise at each $n$ are shown as dashed lines. The PSD knee $\nu_k=\Gamma/\pi$ and amplitude $S_y(0)$ versus $n$ at $\SI{10}{\milli\kelvin}$ for PCR 1, 2 and 3 are shown in (b, c), (d, e) and (f, g), respectively. Fits to scaling laws $\Gamma/\pi = K_1 \times n^a$ and $\mathcal{A}^2/2\Gamma = K_2 \times n^b$ are shown as solid lines. Panels (h) and (i) show the fitted power law exponent values $a$ and $b$ for ten PCRs in the array shown in Fig.~\ref{figure1}a-c. (j) The $Q_i', Q_i$ and $Q_\phi$ of PCR 2 versus $n$ at fixed $T$ = $\SI{10}{\milli\kelvin}$ are shown as blue, green and red circles, respectively. The fit of $Q_\phi$ to $K_3\times n^{0.6}$ is shown a black line.}
    \label{figure4}
\end{figure*}
To analyze how temperature affects the frequency noise, we examine two parameters extracted from $S_y(\nu)$ measurements at fixed phonon number $n$. We show the PSD knee frequency $\Gamma/\pi$ and noise amplitude  $S_y(0)$ in Fig.~\ref{figure3}(a,b), \ref{figure3}(c,d), and \ref{figure3}(e,f) for PCR 1, 2 and 3, respectively. These measurements reveal two transitions occurring at around \qtyrange{100}{200}{\milli\kelvin}. First, the noise amplitude $S_y(0)$ decreases. Secondly, the knee frequency $\Gamma/\pi$ converges with the cavity bandwidth $\kappa/4\pi=(\kappa_e+\kappa_i')/4\pi$ (shown as a grey dot-line). For slow RTS frequency noise $\kappa/4\pi\gg\Gamma_s/\pi$, the expected $S_y(\nu)$ is Lorentzian $S_{\text{RTS}}(\nu) = \mathcal{A}^2/2\Gamma_s \times 1/(1+(\pi\nu/\Gamma_s)^2)$ with a knee $\Gamma_s/\pi$ and $S_y(0)=\mathcal{A}/2\Gamma_s$ with $\mathcal A \equiv A/\bar\omega_0$ \cite{SI}.  If $\Gamma_s/\pi$ approaches or exceeds $\kappa/4\pi$, the measured frequency fluctuations are averaged over the cavity response time. The measured $S_y(\nu)$ is filtered:
\begin{align}
    S_{\text{filt}}(\nu) = \frac{\mathcal{A}^2}{2\Gamma_s}\frac{1}{1+\Big(\frac{4\pi\nu}{\kappa}\Big)^2}\frac{1}{1+\Big(\frac{\pi\nu}{\Gamma_s}\Big)^2}+S_0(\nu) \label{eq:Sfilt}
\end{align}
We derive equation~(\ref{eq:Sfilt}) (see~\cite{SI}) and verify the behavior over the relevant parameter regimes by numerically integrating the resonator time evolution equation and simulating the measured signal. In our model, although the measured PSD knee frequency $\Gamma/\pi$ is limited by the cavity bandwidth $\kappa/4\pi$ at high temperatures, the PSD amplitude $S_y(0)=\mathcal{A}^2/2\Gamma_s$ directly reflects the underlying switching rate $\Gamma_s$ even when it exceeds the cavity bandwidth. Assuming a temperature-independent jump amplitude $\mathcal{A}_0$, we fit the inferred switching rate to a polynomial $\hat\Gamma_s/2\pi=\alpha+\theta T +\beta T^2$ leading to good fits of $S_y(0)$ across all measured temperatures (red lines in Fig.~\ref{figure3}b,d,f). The fits predict that the constant and quadratic terms dominate, and that there is no required linear dependence, \text{i.e.}, $\theta=0$. Moreover, combined with the separately measured $\kappa(T)$, our model of $\hat\Gamma_s(T)/2\pi$ predicts an $S_{\text{filt}}$ knee frequency $\hat\Gamma/\pi$ in line with experiment \cite{SI}, shown as the colored lines in Fig.~\ref{figure3}(a,c,e). We therefore conclude that while the apparent frequency noise $S_y(\nu)$ decreases with temperature, this is due to the switching rate exceeding what the cavity bandwidth can track. The underlying mechanical dephasing time $T_{2,m}^*$, determined by the jump amplitude ($1/T_{2,m}^*\sim\mathcal{A}_0^2$), appears to remain unchanged with temperature. It is likely that the slower switching noise observed at $\SI{10}{\milli\kelvin}$ will be easier to circumvent in future experiments.

\section*{Drive Power Scaling of Frequency Noise}
To determine whether the switching is caused by continuous driving, we vary the incident RF power while at a fixed temperature of $\SI{10}{\milli\kelvin}$. Fig.~\ref{figure4}a shows $S_y(\nu)$ (dot-lines) versus the inferred intra-cavity phonon occupation $n$ with the corresponding detection noise (dashed lines) for PCR 1. We fit each $S_y(\nu)$ to equation~(\ref{eq:Sy}) (solid lines). The PSD parameters $(\Gamma/\pi, \mathcal{A}^2/2\Gamma)$ for PCR 1, 2, and 3 are shown in Fig.~\ref{figure4}(b, c), \ref{figure4}(d, e) and \ref{figure4}(f, g), respectively. Fits to scaling laws $\Gamma/\pi \sim n^a$ and $\mathcal{A}^2/2\Gamma \sim n^b$ are shown as solid lines. Across all PCRs, we typically observe $\Gamma/\pi\sim n^0$ and $\mathcal{A}^2/2\Gamma\sim n^{-1}$, as summarized by Fig.~\ref{figure4}(h,i). In Fig.~\ref{figure4}j we show $Q_i'$, $Q_i$, and $Q_\phi$ as blue, green and red, respectively, of PCR 2 as a function of $n$ at $\SI{10}{\milli\kelvin}$. $Q_\phi$ is fit to $K_3\times n^c$ (black line) yielding $c=0.6$, while $Q_i'$ is power independent. 

\begin{figure}[h]
    \centering
        \includegraphics[width=\linewidth]{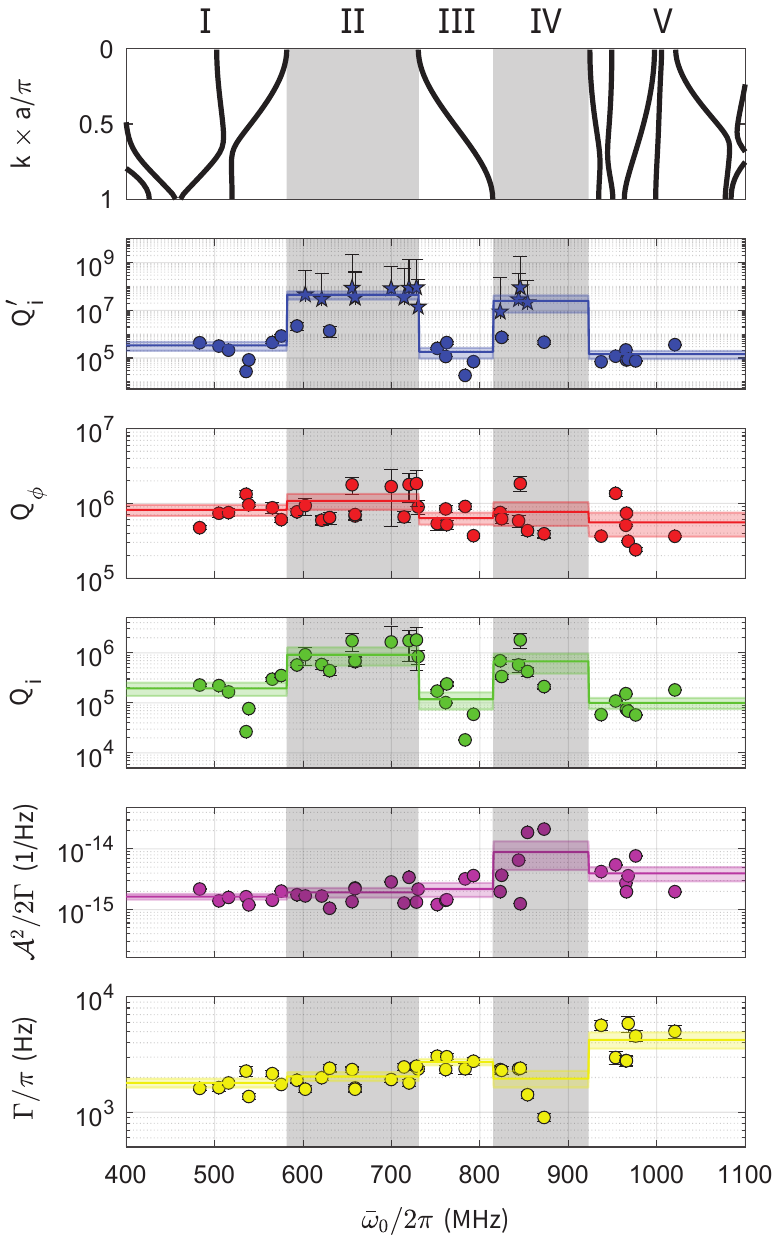}
        \caption{\textbf{Band structure effects on loss and dephasing.} The simulated acoustic band structure for the PCRs in the array (see Fig.~\ref{figure1}a-c) is shown in the top panel. For each resonator in the array, the $Q_i^\prime$, $Q_\phi$, $Q_i$, $\mathcal{A}^2/2\Gamma$ and $\Gamma/\pi$ is shown ($T\approx\SI{10}{\milli\kelvin}$, $n \approx 2\times 10^4$). The solid colored lines are the mean values of the plotted parameter in the frequency bands I, II, III, IV and V, while the colored bands represent one half of a standard deviation. The stars indicate highly overcoupled modes where $\kappa_i^\prime/\kappa_e$ is too small for $\kappa_i'$ to be determined precisely.   
        }
    \label{figure5}
\end{figure}

\section*{Acoustic Band Structure Effects}
TLS properties are strongly affected by engineered phonon environments~\cite{maccabe2020nano,chen2024phonon, odeh2025non}. We therefore study mechanical dephasing and relaxation when the phonon density of states is modified. In Fig.~\ref{figure5} we show $Q_i'$, $Q_\phi$, $Q_i$, $\mathcal{A}^2/2\Gamma$ and $\Gamma/\pi$ of the PCR array in Fig.~\ref{figure1}a-c acquired at \SI{10}{\milli\kelvin} with $n \approx 2\times 10^4$. The size of the phononic crystal defect is scaled between PCRs to enable frequency multiplexing. The top panel of Fig.~\ref{figure5} depicts the finite element simulated acoustic band structure of the periodic phononic crystal cells, which are identically designed for each PCR. The bandgap improves the mean $Q_i'$ by over 100x since radiative relaxation of the mode and the resonant TLS into the substrate is mitigated \cite{chen2024phonon, odeh2025non}. The longest-lived PCRs (marked by stars) have $\kappa_i'/\kappa_e<0.01$, which prevents precise determination of $Q_i'$ via the LCM and ringdown measurements. The stars are the values of $Q_i'$ attained via the LCM while the upper bars indicate the possible range of $Q_i'$ based on the uncertainty in the $S_{11}$ locus diameter. We estimate that $Q_i'$ exceeds 10 million for the best bandgap-protected modes. Unlike $Q_i'$, the overall dephasing $Q_\phi$ (and $\{\mathcal{A}^2/2\Gamma, \Gamma/\pi\}$) is insensitive to the band structure, suggesting that frequency fluctuations are not caused by the mode or resonant TLS interacting with the substrate. For the highest $Q_i'$ PCRs,  $\kappa_\phi$ is $\sim$100x greater than the intrinsic relaxation rate $\kappa_i'$, highlighting how dephasing dominates in acoustically shielded nanomechanical devices at millikelvin temperatures~\cite{ maccabe2020nano, kalaee2019quantum, bozkurt2023quantum, cleland2024studying, bozkurt2024mechanical}.

\section*{TLS-induced Phonon Decoherence} 
 The observed switching behavior suggests a role for TLS in explaining the measured frequency noise ~\cite{phillips1987two, behunin2016dimensional, emser2024thinfilmquartzhighcoherencepiezoelectric}. The TLS in our system can be thought of as two nearly degenerate configurations of atoms in the material constituting the device. Such an atomic complex can couple to vibrations. Meanwhile, our sub-micron-scale mechanical resonator supports a long-lived strain field that concentrates the acoustic energy in a small volume. As such, we expect to operate in a regime where we have strong interaction with a few TLS, rather than the usual regime of weak interactions with many \cite{SI}. This regime is analogous to what's observed in superconducting transmon qubits, where individual TLS in the Josephson junction oxide layer can strongly couple to the qubit. In those systems, researchers have directly observed coherent interactions between single TLS and qubits, leading to phenomena like avoided crossings and vacuum Rabi splitting~\cite{simmonds2004decoherence,martinis2005decoherence,grabovskij2012strain, muller2019towards,chen2024phonon}. Nanomechanical resonators, with their concentrated strain fields,  offer a similar platform for studying and controlling these individual quantum defects, rather than the ensemble effects typically observed in larger mechanical systems. Indeed, similar devices measured in a few-phonon regime exhibited multi-exponential energy decay, which was attributed to coupling to a handful of TLS with slow energy decay~\cite{cleland2024studying}. Moreover, in related work currently under preparation~\cite{Mert2025}, we bias the PCR electrodes to induce a DC strain via the piezoelectric effect and directly observe strong resonant coupling of a single TLS to a mechanical mode characterized by a coupling rate $g_x/2\pi$ on the order of 1~MHz.

Our experiments here show consistent dependencies of frequency noise on temperature, mean phonon number, and band structure across multiple devices, and allow us to assess specific models of dephasing. They point to the role of a far-detuned TLS in generating the observed RTS noise. The persistence of frequency noise at high phonon numbers ($n \gg 100$), well above the saturation limit for resonant TLS~\cite{wollack2021loss}, combined with the bandgap independence of both the quality factor $Q_\phi$ and switching rate $\Gamma_s$, suggests that the responsible TLS are significantly detuned from the mechanical mode. This off-resonant configuration explains why the switching characteristics remain stable even at high drive powers where resonant TLS effects are suppressed. In this far-detuned setting, the interaction between the mode and the TLS are governed by the dispersive Jaynes-Cummings Hamiltonian~\cite{arrangoiz2019resolving,cleland2024studying,wollack2022quantum, SI}: 
\setlength{\abovedisplayskip}{8.5pt}
\setlength{\belowdisplayskip}{8.5pt}
\begin{align}
    \frac{H}{\hbar} = \Bigg(\bar{\omega}_0 + \frac{g_x^2}{\Delta_{\epsilon,m}}\sigma_z\Bigg)\Bigg(a^\dag a+\frac{1}{2}\Bigg)+\frac{\omega_\epsilon}{2}\sigma_z \label{eq:Hdisp}
\end{align}
where $\Delta_{\epsilon,m}=\omega_\epsilon - \bar{\omega}_0$, $\omega_\epsilon$ is the TLS frequency, $\sigma_z$ is the TLS psuedo-spin operator, and $a$ ($a^\dag$) is the phonon annihilation (creation) operator. The dispersive limit holds for $g_x/\Delta_{\epsilon,m}\ll 1$ and $n\ll n_{c,disp}=\Delta_{\epsilon,m}^2/4g_x^2$, which precludes excitation exchange between the TLS and mechanical mode. This model could by itself predict the switching behavior if thermal excitation of the TLS is sufficiently large, i.e., the TLS transition frequency is sufficiently small ($\hbar \omega_\epsilon \ll k_\mathrm{B} T$), to cause the TLS to spend roughly half its time in each state.  For our observed frequency shifts of $A/2\pi \approx 2$ kHz, this model would require a TLS with frequency $\omega_\epsilon/2\pi \approx 50$ MHz (to be thermally excited at 10 mK) and coupling strength $g_x/2\pi \approx 0.5$ to $1$ MHz to a 750 MHz mechanical mode.  Alternatively, strong coupling between a more nearly resonant TLS primarily in its ground state and a much lower frequency two-level fluctuator (TLF), can cause TLS frequency to fluctuate and lead to mechanical frequency noise. The generalized tunneling model (GTM) ~\cite{faoroIoffe2015interacting} invokes TLS-TLF coupling to explain $1/f$ noise from ensembles of near-resonant TLS coupled to thermal fluctuators (TLFs) ~\cite{gao2007noise, gao2008experimental,gao2008physics, gao2008semiempirical, gao2011strongly, kumar2008temperature,  burnett2014evidence, burnett2016analysis,  lucas2023quantum, CorrDecoherence2019Ustinov}. Our observations suggest a simpler scenario: a single far-detuned TLS strongly coupled to a TLF via $H_{\text{TLS-TLF}}/\hbar=(J_z/4) \sigma_z \sigma_z^{th}$, where $\sigma_z^{th}$ is the thermal fluctuator spin operator. Using realistic parameters for this minimal model ($g_x/2\pi \approx 0.5$ MHz, $\Delta_{\epsilon,m}/2\pi \approx 300$ MHz, and $J_z/2\pi \approx 400$ MHz ~\cite{lisenfeld2015observation}), we obtain temperature-independent frequency shifts of $A/2\pi \approx 2$ kHz, consistent with our measurements (see equation~\ref{eq:Amech}). The large detuning ($g_x/\Delta_{\epsilon,m} \approx 1/600 \ll 1$) and $n < n_{c,disp}\approx 2\times 10^5$ ensures operation well within the dispersive regime while maintaining sufficient coupling to produce the observed frequency shifts. The TLS-TLF pair with the largest coupling will dominate the noise. While this simplified picture captures many aspects of our observations, the $T^2$ dependence of the switching rate $\Gamma_s$ deviates from standard predictions for thermal bath coupling ~\cite{faoroIoffe2015interacting,behunin2016dimensional, SI}. This unusual temperature dependence, particularly the saturation at low temperatures, may be due to quasiparticle-mediated TLS switching, possibly in the Al oxide on the electrodes. While poor device thermalization could potentially explain the saturation behavior, other TLS-dependent device properties, such as intrinsic mechanical quality factors and frequency shifts, respond to temperature change even when at base temperature, suggesting the $T^2$ dependence reflects an intrinsic physical process. Similar effects were observed in single TLS coupled to qubits~\cite{chen2024phonon}. However, further experiments are needed to understand this mechanism fully.

\section*{Conclusions} 
In this work, we demonstrate that decoherence in nanoscale phononic crystal resonators is dominated by coupling to TLS. We propose that this noise arises from the strong interaction with a single far-detuned two-level system, which may be thermally excited or strongly coupled to a thermal fluctuator. 

Our measurement approach is based on direct microwave reflectometry of the devices, which lets us measure many modes with different frequencies in parallel. Through a systematic study of temperature, acoustic frequency placement with respect to the phononic bandgap, and drive power dependence, we make several key observations. First, we see a 100-fold reduction in frequency noise as we increase temperature from 10 millikelvin to 1 kelvin, but show that this improvement is due to the switching rate exceeding the cavity bandwidth as opposed to improved coherence. Despite the obvious effect of the phononic bandgap on the intrinsic quality factors, leading to $Q'_i>10^7$, we see that spectral diffusion is unaffected -- consistent with far-detuned defects. Finally we see that dephasing persists at large intra-cavity phonon numbers ($>10^5$) which lends further support to the far-detuned TLS model.

Our work provides insights into the performance limits of nanomechanical devices and points to promising avenues toward noise mitigation for application in quantum science. In quantum information processing, the highly unequal rates for decay and dephasing that we have measured suggest that future work on error correction and dynamical decoupling taking into account the biased nature of the noise may be able to achieve improved performance. In the context of sensing, in particular mass spectrometry, for which these resonators are particularly well suited, our characterization of the frequency stability provides insights on what strategies may enhance mass resolution. In addition to improvements in materials processing, future work will focus on improving measurement and mitigation strategies for TLS, and the development of sensing protocols capable of achieving high mass resolution in the presence of this form of structured noise.


\section*{Methods} \label{Methods}
\subsection{Device Fabrication}
We followed a fabrication process derived from~\cite{wollack2021loss,lee2023strong,cleland2024studying} excluding the superconducting qubit integration. Our phononic crystal resonator (PCR) fabrication begins with 500 nm thick X-cut congruent lithium niobate (LN) bonded to a high resistivity $\langle 111 \rangle$ silicon handle (525~$\mu$m thick, $>$ 3 k$\Omega$-cm). The chip is annealed at 500~C to induce atomic terracing~\cite{gruenke2024surface}. Then the LN is thinned to 250 $\pm$ 10 nm by blanket Ar ion milling (Intlvac Nanoquest) and the thickness is measured via ellipsometry. The PCR patterns, with the phononic crystals along the LN crystal y-axis, are defined via electron beam lithography (JEOL
JBX-6300FS, 100 keV) in a hydrogen silsesquioxane (HSQ, FOx-16) mask and transferred to the LN by angled Ar ion milling. Resist residue and amorphous LN is removed by cleaning with baths of buffered oxide etchant, heated dilute hydrofluoric acid, and piranha. The nanowire electrodes and contact pads are then defined via separate electron beam lithography and photolithography (Heidelberg MLA150, 405 nm) lift-off masks. The 50 nm thick and 300 nm thick 5N aluminum films for the nanowires and contact pads, respectively, are deposited using a Plassys electron beam evaporator (MEB550S). The lift-off masks are cleaned with oxygen plasma to remove organic residue before depositing metal. The bandaging process is enabled by Ar ion milling the Al oxide of the nanowire metal layer in the Plassys immediately before depositing the Al contact pads. Lastly, the nanomechanical devices are released by a timed $\text{XeF}_2$ gaseous dry etch (Xactix). The device packaging consists of an LN chip glued and Al wire bonded to a dielectric circuit board (Hughes). In future work, reducing TLS density could be achieved by annealing after ion milling or by moving the metal off-defect ~\cite{wollack2021loss}. Employing a mechanical mode with lower surface strain may reduce $g_x$ and mitigate decoherence. The dimensions and parameters of the devices in this study are shown in Supplementary Tables~\ref{table:DevDimensions} and \ref{table:DevParams}, respectively. The device geometry and the measurement setup are displayed in Supplementary Figure~\ref{fig:S1} and Supplementary Figure~\ref{fig:S2}.
\newpage

\section*{Acknowledgements}

The authors would like to acknowledge numerous valuable discussions with Prof. M. I. Dykman, K. K. S. Multani, A. Y. Cleland, T. Makihara, Y. Guo and R. G. Gruenke for useful discussions and assistance during fabrication. Part of this work was performed at the Stanford Nano Shared Facilities (SNSF) and at the Stanford Nanofabrication Facility (SNF), supported by the National Science Foundation under award ECCS-2026822. Work was performed in part in the nano@Stanford labs, which are supported by the National Science Foundation as part of the National Nanotechnology Coordinated Infrastructure under award ECCS-1542152.

We gratefully acknowledge support from multiple sources that made this work possible. This material is based upon work supported by the Air Force Office of Scientific Research and the Office of Naval Research under award number FA9550-23-1-0338. Additional funding was provided by Amazon Web Services Inc. We thank the Gordon and Betty Moore Foundation for support through a Moore Inventor Fellowship. This research was also supported by the National Science Foundation CAREER award No. ECCS-1941826. Some of this work was funded by the US Department of Energy through grant no. DE-AC02-76SF00515 and via the Q-NEXT Center.

\section*{Competing interests}
A.H.S.-N. is an Amazon Scholar. The other authors declare no competing interests.

\bibliographystyle{naturemag}
\bibliography{References}

\clearpage
\newpage
\onecolumngrid

\setcounter{section}{0}
\setcounter{figure}{0}
\setcounter{table}{0}
\setcounter{equation}{0}

\section*{Supplementary information}

\renewcommand{\figurename}{Supplementary Figure}
\renewcommand{\thesection}{Supplementary Note \arabic{section}}
\renewcommand{\thefigure}{\arabic{figure}}
\renewcommand{\tablename}{Supplementary Table}
\renewcommand \theequation{S\arabic{equation}}
\setlength{\abovedisplayskip}{9.5pt}
\setlength{\belowdisplayskip}{9.5pt}

\section{Device Details}
\begin{figure*}
    \centering
        \includegraphics[width=\linewidth]{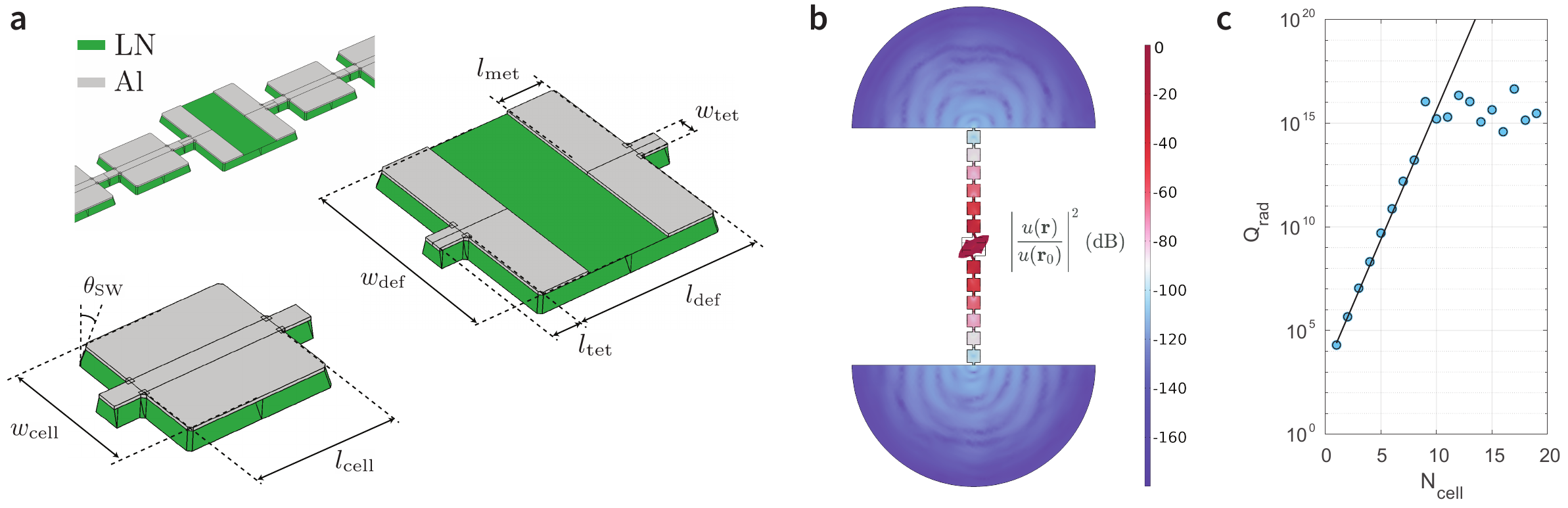}
        \caption{\textbf{Device geometry and its acoustic mode.} (a) The dimensions of a phononic crystal resonator are defined. (b) A finite element simulation of the mechanical deformation for a device with a frequency situated in the bandgap. Radiation into the anchors is suppressed by $>$ 100 dB. (c) Simulated radiative quality factor $Q_{\text{rad}}$ as a function of the number of cells $N_{\text{cell}}$ constituting the phononic crystal. A $Q_{\text{rad}}$ improvement of 12.5 dB per cell (see black line) is in agreement with \cite{maccabe2020nano, bozkurt2023quantum}.}
    \label{fig:S1}
\end{figure*}

\noindent The dimensions and parameters of the devices in this study are summarized in Supplementary Tables~\ref{table:DevDimensions} and~\ref{table:DevParams}, respectively. Supplementary Figure~\ref{fig:S1} specifies the dimensions of a phononic crystal resonator and shows its mechanical shear mode profile which has its frequency placed in the complete acoustic bandgap.

\section{Measurement Setup}

\noindent The full experimental setup is shown in Supplementary Figure~\ref{fig:S2}. Output signal amplification is done using a Low Noise Factory high electron mobility transistor (HEMT) thermalized to the 3K plate followed by low noise amplifiers (LNAs) situated at room temperature. To attenuate standing wave cavities between the amplifiers and block spurious tones, we employ attenuators, extensive filtering and DC blocks between each amplifier. We generally employ Mini-Circuits components to manipulate microwave frequency signals: DC blocks (BLK-18-S+), band-pass filters (BPFs; ZX75BP-750-5+), high-pass filters (HPFs; SHP-150+), low-pass filters (LPFs; VLFX-1050+), Bias tee (ZFBT-6GW+), and Attenuators (FW-3+, VAT-5+, VAT-10+, FW20+). We also employed: LNA1 (Pasternack, PE15A1013), LNA2 (narda Miteq, LNA-30-00101200-17-10p), Diplexer (Marki, DPXN2), and cryogenic attenuators (Bluefors). An RC low-pass filter (667 $\Omega$, 10 nF) at 3K and a 12 k$\Omega$ cryogenic breakout are used for DC biasing the device~\cite{Mert2025}. The mixing chamber (MXC) plate temperature is tracked by a Lakeshore Model 372 AC Resistance Bridge and Temperature PID loop controller. The resistive heater on the MXC is supplied by Bluefors and is driven by a Stanford Research Systems SIM 928 isolated voltage source through a 120 $\Omega$ channel.
 \newline 
 
 \noindent In our experiments, we avoid using a microwave readout cavity or qubit to ensure that we are probing the intrinsic properties of the mechanical system. For instance, these microwave components may have their own frequency fluctuations, hybridize microwaves with the phonons, or exhibit quasiparticle-induced noise/nonlinearity at temperatures close to the critical temperature. In future work, we would employ a broadband Josephson junction-based quantum-limited amplifier at 10 mK (e.g. \cite{macklin2015near}) as the first amplifier in the chain to maximize the signal-to-noise ratio to study even faster timescale fluctuation processes. This would require higher frequency devices situated within the gain band.

\section{Random Telegraph Noise} \label{sec:RTS}
\noindent Suppose we have a zero-mean random telegraph signal (RTS), $x(t)$, that spends equal time between two values, $-A$ and $+A$, with a switching rate of $\Gamma_{s}$. The probability of $m$ switches in time interval $\tau$ is given by the Poissonian distribution  $P_m(\tau) = (\Gamma_{s} \tau)^m e^{-\Gamma_{s} \tau}/m!$. Since an \{even, odd\} number of switches in time $\tau$ implies $\{x(t)x(t+\tau) = A^2$,  $x(t)x(t+\tau) = -A^2$\}, the autocorrelation function of the signal is:
\begin{align}
\langle x(t)x(t+\tau) \rangle &= A^2 (P_0(\tau)+P_2(\tau)+\cdots) - A^2 (P_1(\tau)+P_3(\tau)+\cdots) \nonumber \\
 &= A^2 (e^{-\Gamma_{s} \tau} + \frac{(\Gamma_{s} \tau)^2}{2!} e^{-\Gamma_{s} \tau} + \cdots) - A^2 (\Gamma_{s} \tau e^{-\Gamma_{s} \tau} + \frac{(\Gamma_{s} \tau)^3}{3!} e^{-\Gamma_{s} \tau} + \cdots) \nonumber \\
 &= A^2 (\cosh (\Gamma_{s} \tau) -  \sinh (\Gamma_{s} \tau))e^{-\Gamma_{s} \tau} = A^2 e^{-2\Gamma_{s} \tau}
\end{align}
We calculate the power spectral density (PSD) using the Wiener-Khinchin theorem:
\begin{align}
S_{\text{RTS}}(\Omega) =  \frac{A^2}{2\Gamma_s}\left( \frac{1{}}{1 + \left(\frac{\Omega}{2\Gamma_s}\right)^2} \right) \label{eq:SyRTS}
\end{align}
The knee (3 dB roll-off point) is then $\Omega_{\text{k}}= 2\Gamma_{s}$, or $\Omega_k/2\pi = \nu_{\text{k}}=\Gamma_{s}/\pi$ and $\tau_0=1/2\Gamma_s$ is the correlation time.

\newpage
\section{General Behavior of a Fluctuating Resonator}

\noindent We wish to study a linear resonator in the presence of a classical fixed frequency drive $a_{in}(t)=\alpha_{in}e^{-i\omega t}$. The relevant Hamiltonian is:
$H/\hbar = \omega_0(t)a^\dag a+ \sqrt{\kappa_e}(\alpha_{in}a^\dag e^{-i\omega t} + \text{h.c.})$, where $\kappa_e$ is the extrinsic coupling rate assumed to be constant and $\omega_0(t) = \omega_0+\delta(t)$ is the resonant frequency. In the rotating frame of the drive, the Langevin equation of motion for the intra-cavity field $a(t)$ is:
\begin{align}
    \frac{d}{dt} a(t) = \left(-i\Delta-i\delta(t)- \frac{\kappa}{2} - \frac{\theta(t)}{2} \right)a(t)-\sqrt{\kappa_e}\alpha_{in} \label{eq:fullEOM}
\end{align}
where $\delta(t)$ and $\theta(t)$ are the fluctuations in frequency and intrinsic linewidth, respectively, while $\Delta=\omega_0-\omega$ is the drive detuning, $\kappa$ is the total linewidth and $a$ is the intra-cavity field operator. We assume that the fluctuations are much slower than the drive frequency, resonant frequency, and the analog-to-digital sampling frequency of the measurement instrument, which are all on the order of 1 GHz.

\subsection{Steady-State Behavior}
\noindent First, let’s assume that 1) the integration time $\tau=0$ and 2) the fluctuation timescales are much slower than $\kappa/2$. The cavity field has sufficient time to respond to a frequency or linewidth change, ensuring that $da/dt=0$. At any point in time we have a quasi steady-state solution to equation \ref{eq:fullEOM}:

\begin{align}
  a(t)=\frac{-\sqrt{\kappa_e}}{i(\Delta+\delta(t))+(\kappa+\theta(t))/2}a_{in}  
\end{align}
Using the symplectic boundary condition relating the input $a_{in}$, output $a_{out}$ and intra-cavity $a$ fields of \cite{collett1985input}:
\begin{align}
    a_{out}(t) = a_{in}(t)+\sqrt{\kappa_e} a(t)
\end{align}
we find:
\begin{align}
a_{out}(t)=\left(1-\frac{{\kappa_e}}{i(\Delta+\delta(t))+(\kappa+\theta(t))/2}\right)a_{in}=S_{11}(\delta(t),\theta(t))a_{in}
\end{align}
The expression for $S_{11}$ after adding the cavity asymmetry angle $\phi$ and microwave background $S_{\text{bg}}$ is equation 1 in the main text:
\begin{align}
	S_{11}(\omega, t) = I(\omega,t)+iQ(\omega,t) = S_\mathrm{\text{bg}}\times\Bigg(1-e^{i\phi}\frac{\kappa_e/\cos(\phi)}{i\Delta(t)+\frac{\kappa(t)}{2}}\Bigg), 
\end{align}
where $S_{\text{bg}}=\sum_{j=0}^na_j\Delta^j\exp[i\sum_{k=0}^nb_k\Delta^k]$ is a complex polynomial background fit, $\Delta(t)=\Delta+\delta(t)$, and $\kappa(t)=\kappa+\theta(t)$. In our experiment, we measure the reflection parameter $S_{11}(t)$, from which we can extract $\delta(t)$ and $\theta(t)$. If the cavity is undergoing slow RTS frequency fluctuations, the inferred PSD will be described by equation \ref{eq:SyRTS}. This slow fluctuation regime is the typical regime of study of parameter noise in superconducting circuits, resonators and micro/nanomechanical devices. 

 \newpage
\subsection{Cavity-Based Filtering}
\noindent In general, the reflection signal does not capture the true frequency fluctuations due to the finite cavity linewidth. We perform a perturbative analysis to understand how a finite cavity bandwidth alters the measured noise spectral density. First, we decompose the intra-cavity field into steady-state and fluctuating parts: $a(t) = a_0+\delta a(t)$ and assume a resonant drive and small fluctuations compared to $\kappa$. Without fluctuations $\delta=\theta=0$ we find $a_0=-2\sqrt{\kappa_e}a_{in}/\kappa$. Substituting $a(t) = a_0+\delta a(t)$ into equation \ref{eq:fullEOM} and ignoring second order terms (e.g. $\delta(t)\delta a(t)$) we find:
\begin{align}
    \frac{d (\delta a(t))}{dt} = -\frac{\kappa}{2}\delta a(t) - \Bigg(i\delta(t)+\frac{\theta(t)}{2}\Bigg)a_0 \label{eq:linearDeltaa}
\end{align}
We aim to find the estimated frequency fluctuations $\hat\delta(t)=-(\kappa^2/4\kappa_e)\Im[\delta S_{11}(t)] = (\kappa^2/4\sqrt{\kappa_e}a_{in}) \times \Im[\delta a(t)]$. This requires solving the linear equation \ref{eq:linearDeltaa} for $\delta a$, which can be done by moving into the Fourier domain:
\begin{align}
    \delta a(\Omega) = \frac{-i\delta(\Omega)-\frac{\theta(\Omega)}{2}}{-i\Omega+\frac{\kappa}{2}}
\end{align}
Using the the Fourier transform property $\mathcal{F}\{\delta a^*(t)\}=(\delta a(-\Omega))^\ast$ and assuming $\delta(t)$ and $\theta(t)$ are real functions, we arrive at:
\begin{align}
    S_{\hat\delta}(\Omega) = \frac{S_\delta(\Omega)}{1+\Big(\frac{\Omega}{\kappa/2}\Big)^2}
\end{align}
That is, the estimated frequency fluctuation spectral density $S_{\hat\delta}(\Omega) = |\hat\delta(\Omega)|^2$ is given by the true frequency spectral density $S_\delta(\Omega) = |\delta(\Omega)|^2$ filtered by the cavity response, which has a knee at $\Omega_k = \kappa/2$. If the real frequency noise is described by a telegraph signal, $S_\delta(\Omega)=S_{\text{RTS}}(\Omega)$,
we denote $S_{\hat\delta}(\Omega)\equiv S_{\text{filt}}(\Omega)$ (see equation 6, main text) and its knee frequency is:
\begin{align}
\frac{\hat\Gamma}{\pi} = \frac{1}{\sqrt{2}}\Bigg[\sqrt{\Big(\frac{\kappa}{4\pi}\Big)^4+\Big(\frac{\Gamma_s}{\pi}\Big)^4+6\Big(\frac{\kappa}{4\pi}\Big)^2\Big(\frac{\Gamma_s}{\pi}\Big)^2} -\Big(\frac{\kappa}{4\pi}\Big)^2-\Big(\frac{\Gamma_s}{\pi}\Big)^2\Bigg]^{1/2} \label{eq:KneeAnalyic}
\end{align}
Using the measured cavity knee $\kappa/4\pi$ and the switch rate model $\hat\Gamma_s/\pi=\alpha+\beta T^2$, we compute the expected PSD knee $\hat\Gamma/\pi$ with equation \ref{eq:KneeAnalyic} shown as a light blue line in Fig. 3 (a), (c) and (e) in the main text. This model shows good agreement with the measured knee $\Gamma/\pi$, indicating that equation $S_{\text{filt}}(\Omega)$ and $\hat\Gamma_s$ explains our data well. The spectral density may also be filtered by the integration bandwidth: $1/[1+(\nu/f_c)^2]$ specified by a cutoff frequency $f_c$. In our experiments, $f_c$ = 25 kHz is greater than the cavity bandwidth $\kappa/4\pi\sim$ 10 kHz and the RTS knee 1 kHz $<\Gamma_s/\pi<$ 20 kHz. We therefore see no integration bandwidth filtering that affects our measured PSDs.

\newpage
\section{Frequency Fluctuations from Individual Atomic-Scale Defects}
\setlength{\abovedisplayskip}{6.5pt}
\setlength{\belowdisplayskip}{6.5pt}

\noindent The disordered Al-LN interface, Al oxide, and the LN near the surface damaged by ion milling can host defects, which are generally described as two nearly degenerate configurations of atoms. The standard tunneling model (STM)  ~\cite{phillips1987two, behunin2016dimensional, gao2008physics, muller2019towards,emser2024thinfilmquartzhighcoherencepiezoelectric} treats a two-level system (TLS) defect as a particle in an asymmetric double-well potential with energy:
\begin{align}
H_{\text{TLS}} = \frac{1}{2}(\Delta_0\Sigma_x +\Delta_{as}\Sigma_z)
\end{align}
where $\Delta_0 (\Delta_{as})$ is the tunneling (asymmetry) energy and $\Sigma_x (\Sigma_z)$ is the transverse (longitudinal) psuedo-spin operator for the TLS in the left-right basis. We find the TLS eigenstates by moving into the diagonal basis using the transformation \cite{phillips1987two,Clerk2021PositiveFreqNoise}:
\begin{align}
    \Sigma_z=\cos(\varphi)\sigma_z+\sin(\varphi)\sigma_x \label{diagSz} \\
    \Sigma_x=\sin(\varphi)\sigma_z-\cos(\varphi)\sigma_x \label{diagSx}
\end{align}
where $\tan(\varphi)=\Delta_0/\Delta_{as}$ to find $H_{\text{TLS}} = \epsilon\sigma_z/2$. The energy separating the ground and excited eigenstates is $\epsilon = \hbar\omega_{\epsilon}=\sqrt{\Delta_0^2+\Delta_{as}^2}$. A strain field $\xi$ can perturb the local atomic environment, represented to first order as $\Delta_{as}\to\Delta_{as}+2\gamma:\xi$ where $\gamma\equiv(1/2)\times\partial\Delta_{as}/\partial\xi$ is the elastic dipole moment of the TLS and $:$ denotes the tensor contraction. The change in the system energy takes the form of a dipole interaction, $H_{\text{int}}=(\gamma:\xi)\Sigma_{z}$, which is expressed in the diagonal basis as \cite{behunin2016dimensional}: 
\begin{align}
H_{\text{int}} = \Big[\Big(\Delta_0/\epsilon\Big)\sigma_x + \Big(\Delta_{as}/\epsilon\Big)\sigma_z\Big]\gamma:\xi\equiv\hbar(g_x\sigma_x+g_z\sigma_z)(a+a^\dag)
\end{align}
where $a$ ($a^\dag$) is the phonon annihilation (creation) operator. Assuming there is no phonon population in the solid, we have $\xi(\bold{r}) = \xi_{zp}(\bold{r})(a+a^\dag)$ as the vacuum strain field at the location of the TLS. The vacuum transverse coupling rate responsible for phonon exchange between the strain field and TLS is therefore:
\begin{align}
g_x \equiv (1/\hbar)(\Delta_0/\epsilon)\gamma:\xi_{zp}(\bold{r}) \label{eq:gx}
\end{align}
while the vacuum longitudinal coupling rate describing the dephasing of the TLS caused by the strain is
\begin{align}
g_z \equiv (1/\hbar)(\Delta_{as}/\epsilon)\gamma:\xi_{zp}(\bold{r}) \label{eq:gz}
\end{align}
The full Hamiltonian describing a driven mechanical resonator coupled to a TLS is:
\setlength{\abovedisplayskip}{8pt}
\setlength{\belowdisplayskip}{8pt}
\begin{align}
    \frac{H}{\hbar} &= \omega_0a^\dagger a + \omega_\epsilon \sigma_+ \sigma_- + (g_x\sigma_x+g_z\sigma_z) \left( a + a^\dagger \right) + \sqrt{\kappa_e} \left( a^\dag\alpha_{\text{in}}e^{-i\omega t} + \text{h.c.}\right)
\end{align}
where $\sigma_\pm=\sigma_x\pm i\sigma_y$ are the raising and lowering operators for the TLS. To remove the time dependence of the drive, we move into the interaction frame by applying the unitary transformation $U = \text{exp}[-i ( a^\dag a + \sigma_+\sigma_-)\omega t]$. After applying the rotating wave approximation we arrive at the Jaynes-Cummings (JC) Hamiltonian:
\begin{align}
	\frac{H'}{\hbar} &= \Delta a^\dagger a + \Delta_{\epsilon} \sigma_+ \sigma_- + g_x \left( a \sigma_+ + a^\dagger \sigma_- \right) + \sqrt{\kappa_e} \alpha_{\text{in}} \left( a + a^\dagger \right)
	\label{HamiltonianJC} 
\end{align}
\noindent
where \(\Delta=\omega_0-\omega\) and \(\Delta_{\epsilon}=\omega_{\epsilon}-\omega\). 
\newline

\noindent A key assumption of STM is that TLS have a large distribution of $\Delta_{as}$ and form a quasicontinuum bath. Many TLS are assumed to interact with the resonator at any given time. However, for our nanomechanical device, assuming that TLS reside within a 5 nm thick  damaged LN surface, a 5 nm thick native Al oxide or a 5 nm thick LN-Al interface, the total host volume is $V_h\approx 0.1\text{ }\mu\text{m}^3$. Assuming a density of TLS typical for disordered solids $P_0=1\times10^{44}\text{/Jm}^3$ \cite{black1978relationship, behunin2016dimensional,chen2024phonon}, we find the frequency density of TLS in our device: $\rho_{\text{TLS}}=V_hP_0\approx$ 6.6 TLS/GHz (1 TLS/150 MHz). It is therefore statistically likely that the nearest TLS will be far off-resonant from our $10$ kHz linewidth mode. 
Moreover, we can estimate the transverse coupling rate by simulating the RMS strain in COMSOL, $\xi_{zp}=\sqrt{\hbar\omega_0/2\mathcal{E}V_m}\approx1.5\times10^{-9}$ m/m, where $\mathcal{E}$ is the elastic modulus and $V_m\approx 1\text{ }\mu\text{m}^3$ is the mode volume. Assuming $\Delta_0/\epsilon\approx 1$ and a $\gamma\approx 1$ eV found in silica \cite{golding1976phonon,maccabe2020nano}, we use equation \ref{eq:gx} to find $g_x/2\pi\approx 0.36$ MHz. Over $\sim 50\%$ of the geometry points have a strain at or above the RMS strain, suggesting higher coupling rates are possible. Moreover, $\gamma$ may be higher in LN than silica due to piezoelectricity. In work under preparation \cite{Mert2025}, we tune the frequency of individual TLS in-situ by applying a DC bias across our device electrodes. We measure strong coupling of a PCR mode to individual TLS, with $g_x/2\pi \approx 2\text{ MHz}$ being the highest coupling rate observed. These measurements also confirm the highly discontinuous nature of the TLS frequency spectral density, $\rho_{\text{TLS}}$.
\newpage
\noindent These same TLS when detuned far off-resonance can dominate frequency noise at high powers where resonant TLS effects are power saturated. We are therefore interested in the far detuned (dispersive) regime where the parameter $g_x/\Delta_{\epsilon,m}$ is small ($\Delta_{\epsilon,m} = \omega_\epsilon-\omega_0$) and there is no excitation exchange between the mode and TLS. As such, we apply the unitary transformation $U=\exp[-g_x/\Delta_{\epsilon,m}(\sigma_+a-\sigma_-a^\dag)]$ to equation \ref{HamiltonianJC} and keep the terms up to second order in $g_x/\Delta_{\epsilon,m}$:
\begin{align}
    \frac{H'_{\text{disp}}}{\hbar} = \Bigg(\Delta + \frac{g_x^2}{\Delta_{\epsilon,m}}\sigma_z\Bigg)\Bigg(a^\dag a+\frac{1}{2}\Bigg)+\frac{\Delta_\epsilon}{2}\sigma_z + \sqrt{\kappa_e} \alpha_{\text{in}} \left( a + a^\dagger \right) \label{eq:SHdisp}
\end{align}
which is valid assuming $g_x/\Delta_{\epsilon,m}\ll 1$. If we exclude the drive, we recognize equation \ref{eq:SHdisp} as equation \ref{eq:Hdisp} in the main text. Within the STM, mutual TLS-TLS interactions are ignored. Therefore, fluctuations in the TLS state or frequency can originate solely from a stochastic bath (e.g. thermal phonon or quasiparticle bath coupling). If the TLS hops between its eigenstates, $\sigma_z=\pm1$, at a rate $\Gamma_s$ the mechanical resonator frequency will undergo telegraphic jumps described by the correlator:
\begin{align}\expval{\delta\omega_0(\tau)\delta\omega_0(0)} = A^2e^{-2\Gamma_s\tau} \label{eq:RTScorr}
\end{align}
where $A = 2g_x^2/\Delta_{\epsilon,m}$. We expect symmetric telegraph noise described by equation \ref{eq:SyRTS} and evidenced by Fig. 2 (b). This occurs if the TLS excitation and de-excitation rates are roughly equal, restricting the TLS to low frequencies. A more dominant source of frequency noise could arise if the TLS is strongly dipole coupled to a low frequency two-level fluctuator (TLF) \cite{faoroIoffe2015interacting}. We denote the operators for the TLF with superscript $th$. The TLS-TLF coupling arises as phonons emitted from the TLF can modify the local TLS environment ($\Delta_{as}$) and vice versa. The interaction is described by \cite{lisenfeld2016decoherence}:
\begin{align}
    \frac{H_{\text{TLS-TLF}}}{\hbar}=\frac{J}{4}\Sigma_z\Sigma_z^{th}
\end{align}
The dipole coupling strength is strong if the TLF-TLS spatial separation $R$ is small:
\begin{align}
    J\sim\frac{\gamma:\gamma^{th}}{\rho c_s^2 R^3}
\end{align}
where $\rho$ is the mass density and $c_s$ is the acoustic wave velocity of the solid hosting the TLS-TLF pair. We transform to the diagonal TLF-TLS basis using equations \ref{diagSz} and \ref{diagSx} to find:
\begin{align}
    \frac{H_{\text{TLS-TLF}}}{\hbar}=\frac{J_z}{4}\sigma_z\sigma_z^{th} \label{eq:HTLS-TLF}
\end{align}
where the dressed coupling rate is: $J_z = J\cos(\varphi)\cos(\varphi^{th})$. We ignore the $\sigma_x^{th}\sigma_x$ term which describes resonant TLF-TLS coupling, and terms like $\sigma_x^{th}\sigma_z$ and $\sigma_z^{th}\sigma_x$ that simply offset the TLS and TLF energy levels. In our model the TLF and TLS are far detuned from each other. Equation \ref{eq:HTLS-TLF} indicates that a change in the state of the TLF, $\sigma_{z}^{th}=\pm 1$, will shift the frequency of the TLS between $\omega_\epsilon\pm J_z/2$. If the TLS eigenstate is unchanged during this process, the amplitude of the mechanical frequency jumps will be:
\begin{align}
    A=g_x^2 \Bigg|\frac{1}{\Delta_{\epsilon,m}+J_z/2}-\frac{1}{\Delta_{\epsilon,m}-J_z/2 }\Bigg| \label{eq:Amech} 
\end{align}
and the correlator is the same form as equation \ref{eq:RTScorr} where $\Gamma_s$ is now the TLF switch rate. The noise will be symmetric RTS since the low frequency TLF is expected to be highly thermally populated. In work under preparation \cite{Mert2025}, we directly observe a temperature-independent $g_x$ and $\omega_\epsilon$, suggesting that $A$ may also be temperature-insensitive. This would be consistent with our findings in the main text.

\newpage
\section{Two-Level System Switching via Thermal Bath Coupling}

\noindent Here we examine mechanisms of TLF (or TLS) switching, which we have measured to be $\Gamma_s\sim n^0(\alpha+\beta T^2)$ for all devices. This dependence is unexpected for a TLF, as the primary loss channel is typically assumed to be the thermal radiation bath, described by \cite{maccabe2020nano,behunin2016dimensional, maccabe2020nano}:
\begin{align}
    \frac{H_{\text{TLF-bath}}}{\hbar} = \frac{\omega_{\epsilon}^{th}}{2}\sigma_z^{th} + \sum_q\omega_qb_qb_q^{\dag} + \sum_q (g_{z,q}\sigma_z^{th}+g_{x,q}\sigma_x^{th})(b_q+b_q^{\dag}) 
\end{align}
where $g_{x,q}$ ($g_{z,q}$) is the transverse (longitudinal) coupling of the TLF of frequency $\omega_\epsilon^{th}$ to the the $q^{th}$ thermal bath mode. Ignoring dephasing, the TLF relaxation is dominated by the $g_x$ term which gives rise to a relaxation rate \cite{behunin2016dimensional,faoroIoffe2015interacting, maccabe2020nano}:
\begin{align}
    (\Gamma_1^{th})_{\text{ph, disc}} &= \sum_q \frac{|g_x|^2 \Gamma_q}{[(\omega_\epsilon^{th})^2-\omega_q^2]^2+\omega_q^2\Gamma_q^2}\coth(\hbar\omega_\epsilon^{th}/2k_BT) \nonumber \\ 
    &\approx \sum_q \frac{|g_x|^2 \Gamma_q}{(\omega_\epsilon^{th}-\omega_q)^2+(\Gamma_q/2)^2}\coth(\hbar\omega_\epsilon^{th}/2k_BT)
\end{align}
where the approximation captures bosonic modes near the TLF resonance only, which should  dominate the loss. The $\Gamma_q$ is the decay rate for the $q^{th}$ radiation mode, and $T$ is the temperature of the bath which defines the occupation of each mode via the Bose-Einstein distribution. With a continuum of radiation modes, we replace the sum with an integral that is dominated by frequencies within the range: $\omega_\epsilon^{th}\pm\Gamma_q/2$:
\begin{align}
    (\Gamma_1^{th})_{\text{ph, cont}} \approx 2\pi|g_x(\omega_\epsilon^{th})|^2\rho_{\text{ph}}(\omega_\epsilon^{th})\coth(\hbar\omega_\epsilon^{th}/2k_BT)
\end{align}
where $\rho_{\text{ph}}(\omega)$ is the phonon density of states in the solid hosting the TLF. Regardless of the dimensionality of the bath \cite{behunin2016dimensional}, we inevitably arrive at a function of the form:
\begin{align}    
\Gamma_1^{th}(T)=\gamma_1^{th}\coth(\hbar\omega_\epsilon^{th}/2k_BT)\approx \gamma_1^{th} \times\frac{k_BT}{\hbar\omega_{\epsilon}^{th}}
\end{align}
where $\gamma_1^{th}$ is a temperature independent constant and the linear approximation holds for ${\hbar\omega_\epsilon^{th}\ll k_BT}$. We observe a $\Gamma_s(T)$ that disagrees with this model. We can also rule out a TLF with frequency near the phononic bandgap as the source of frequency noise since we would expect a switch rate dependent on the phonon density of states, $\rho_{\text{ph}}(\omega)$. The TLF would have to be at much lower or higher frequency than those shown in Fig. 5.

\newpage

\begin{table}
\setlength{\tabcolsep}{12pt}
\begin{tabular}{ll}
\toprule
    \textbf{Parameter} & \textbf{Value} \\  \midrule
    LN thickness, $t_{\text{LN}}$ &  250 nm \\
    Al Thickness, $t_{\text{Al}}$ &  50 nm \\
    Defect Width, $w_{\text{def}}$ & 2.14 - 7.62 $\mu$m \\
    Defect Length, $l_{\text{def}}$ & 2.95 $\mu$m \\
    Cell Width, $w_{\text{cell}}$ & 2.13 $\mu$m \\
    Cell Length, $l_{\text{cell}}$ & 2.08 $\mu$m \\
    Tether Width, $w_{\text{tet}}$ & 320 nm \\
    Tether Length, $l_{\text{tet}}$ & 430 nm \\
    Metal Length, $l_{\text{met}}$  & 780 nm \\
    LN Sidewall Angle, $\theta_{\text{SW}}$ & 11$^0$\\
    Corner Radius & 50 nm \\
    \bottomrule

\end{tabular}
\caption{ \textbf{Device dimensions.} The dimensions of the phononic crystals in the array of Fig. 1 estimated from scanning electron microscopy ($w_{\text{def}},l_{\text{def}},w_{\text{cell}},l_{\text{cell}},w_{\text{tet}},l_{\text{tet}}, l_{\text{met}}$) $\pm$ 10 nm and ellipsometry $t_{\text{LN}}$ $\pm$ 10 nm. The $t_{\text{Al}}$ $\pm$ 5 nm is estimated using a quartz crystal monitor in the Plassys during the deposition of Al, while $\theta_{\text{SW}}\text{ }\pm$ $1^o$  is an ion mill parameter. The corner radius was estimated for our finite element simulations.}
\label{table:DevDimensions}
\end{table}

\begin{table}
\setlength{\tabcolsep}{8pt}
    \resizebox{\textwidth}{!}{
\begin{tabular}{cccccccccccc} 
\toprule
    {PCR\#} & {$\bar\omega_0/2\pi$ (MHz)} & {$n/10^4$} & {$\phi$} & {$\kappa_e/2\pi$ (Hz)} & {$\kappa_i/2\pi$ (Hz)} & {$\kappa_\phi/2\pi$ (Hz)} & {$\kappa_i'/2\pi$ (Hz)} & {$\mathcal{A}^2/2\Gamma_s\times10^{15}$ (1/Hz)} & {$\Gamma/\pi$ (Hz)}  \\ 
\midrule
    1 & 751.7 & 2.2 & 0.15 & 8625 & 4377 & 1401 & 2977 & 1.2 & 3058 \\[2mm]
    2 & 761.4 & 2.2 & 0.15 & 8531 & 7522 & 903 & 6620 & 1.4 & 2342 \\[2mm]
    3 & 783.3 & 2.4 & 0.049 & 7933 & 43042 & 861 & 42181 & 3.2 & 2378 \\[2mm]
    4 & 762.6 & 2.4 & 0.18 & 8264 & 3190 & 1467 & 1722 & 1.4 & 3024 \\[2mm]
    5 & 792.8 & 2.9 & 0.28 & 7490 & 13409 & 2145 & 11263 & 3.6 & 2766 \\[2mm] 
    6  & 515.7 & 2.2 & -0.10 & 5665 & 3113 & 685 & 2428 & 1.6 & 1794 \\[2mm]
    7  & 538.7 & 2.2 & 0.021 & 1597 & 7011 & 560 & 6450 & 1.2 & 1368 \\[2mm]
    8  & 565.3 & 2.1 & -0.08 & 6919 & 1902 & 647 & 1255 & 1.4 & 2160 \\[2mm]
    9  & 575.4 & 2.1 & -0.022 & 7436 & 1638 & 950 & 688 & 2.0 & 1740 \\[2mm]
    10 & 976.2 & 2.9 & -0.45 & 16389 & 17004 & 4081 & 12923 & 7.7 & 4570 \\[2mm]
\bottomrule
\end{tabular}}
\caption{\textbf{Device parameters.} Parameters of phononic crystal resonators measured at 10 mK that were studied in Fig. 1 - 4.}
\label{table:DevParams}
\end{table}

\begin{figure*}
    \centering
        \includegraphics[width=0.6\linewidth]{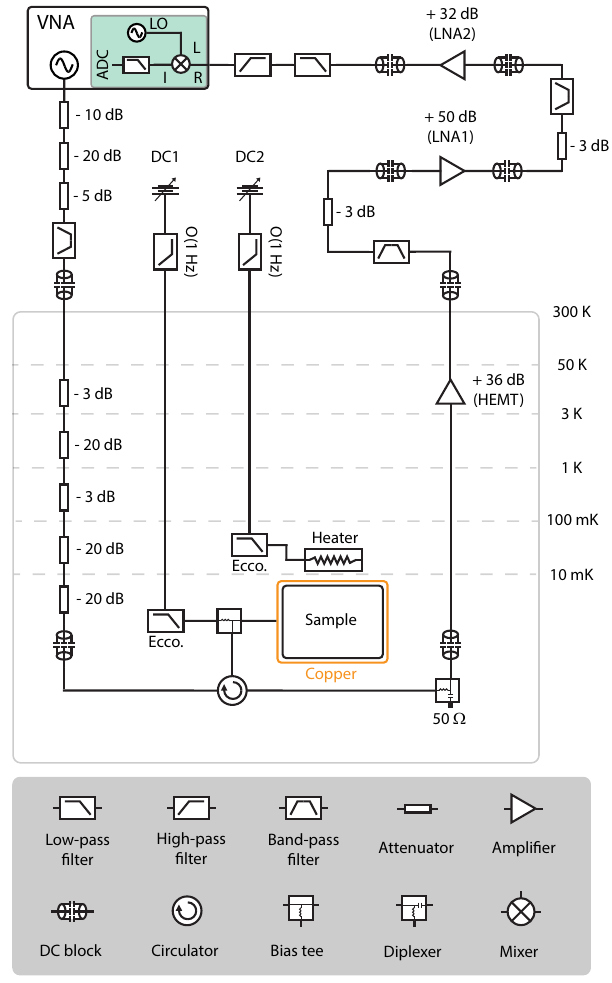}
        \caption{\textbf{Experimental Setup}. The sample is mounted on the mixing-chamber (MXC) plate of a dilution refrigerator (Bluefors, LD250). The chip package is fastened inside an oxygen-free copper enclosure. The vector network analyzer (VNA, Rohde \& Schwarz, ZNB20) sends a microwave tone into the fridge which reflects off the devices via a cryogenic circulator (Quinstar, QCI-M6009001AU) to an NbTi-NbTi superconducting output line. All other lines are made from SCuNi-CuNi. The weak signal is filtered and amplified before demodulation at the VNA. A DC source (DC1, Yokogawa, GS200) maintains the bias across the phononic crystal resonator electrodes at 0 V. Nonzero bias is employed for measurements published elsewhwere \cite{Mert2025}. The second isolated DC source (DC2, Stanford Research Systems, SIM928) applies a DC current through a resistive heater thermalized to the MXC plate, which alters the temperature of the nearby sample. For both DC lines we employ coaxial low-pass infrared filters made from Eccosorb (Ecco.) with 20 GHz cutoffs.}
    \label{fig:S2}
\end{figure*}

\end{document}